\documentstyle[twocolumn,psfig,rotating]{mn}
\def\nn{n}

\def\lsim{~\rlap{$<$}{\lower 1.0ex\hbox{$\sim$}}}
\def\bsim{~\rlap{$>$}{\lower 1.0ex\hbox{$\sim$}}}
\def\sp{{ s}}
\def\ltsim{\lower.5ex\hbox{$\; \buildrel < \over \sim \;$}}
\def\gtsim{\lower.5ex\hbox{$\; \buildrel > \over \sim \;$}}

\def\dd{{\rm d}}

\def\pmb#1{\setbox0=\hbox{#1}%
\kern-.025em\copy0\kern-\wd0
\kern.05em\copy0\kern-\wd0
\kern-.025em\raise.0433em\box0}

\begin{document}
\title[self-similar expansion]
{Self-similar cosmological expansion of collisional gas}
\author[Chuzhoy \& Nusser ]{Leonid Chuzhoy
and
Adi Nusser\\
Physics Department,
Technion, Haifa 32000, Israel\\
E-mail: cleonid@tx.technion.ac.il,  adi@physics.technion.ac.il
}

\maketitle

\begin{abstract}
  We derive similarity solutions for the expansion of negative initial
  density perturbations $\delta M/M \propto r^{-\sp}$ $(\sp>0)$ in an
  Einstein-de Sitter universe filled with collisional baryonic gas and
  collisionless matter.  The solutions are obtained for planar,
  cylindrical, and spherical perturbations.  For steep perturbations,
  central cavities surrounded by over-dense shells expanding as
  $t^{{2(s+1)}/{(3s)}}$ develop in both matter components.  We also
  consider the case when baryonic shells are driven by internal
  pressure.  Before redshift $\sim 8-10$ baryonic shells cool by
inverse Compton scattering with the 
  cosmic background radiation
  and can form bound objects of mass $\sim 10^{11}\Omega_{\rm
  b}^{-1/2} M_{\odot} (1+z)^{-3/2}$.  Hot shocks of radius $R$ cause a
  Compton y-distortion of order $y=10^{-5}\Omega_{\rm b} (R/10 {\rm Mpc})^2
  f_F$, where $f_F$ is the volume filling factor of these shells.
\end{abstract}


\section {Introduction}

Gravitational collapse of matter in initially over-dense regions leads
to the formation of dense bound objects in the universe.  Star
formation is efficient in these objects and so we expect them to
harbor most of the luminous matter in the universe.  In the linear
regime, gravitational instability simply amplifies the density
fluctuations by a universal time dependent factor.  Thus, an initially
under-dense region remains under-dense and cannot lead to the
formation of dense structures.  In the nonlinear regime, however, the
evolution of under-dense regions depends strongly on the profile of
its density perturbation.  In an isolated under-dense region, the
density remains below the background if the profile is shallow.  If
the profile is steep then expanding inner shells can catch up with
shells farther out causing the appearance of caustics in the
collisionless matter component and shocks in the baryonic gas (e.g.,
Fillmore \& Goldreich 1984, Bertschinger 1985).  Caustics and shocks
are very dense and can fragment under their self-gravity to form bound
structures.
 
 In a generic density field, a
large scale under- or over-dense region
 contains small scale substructure of
density fluctuations.  In a large
 scale over-density, substructures merge
under the large scale
 gravitational field and become part of a large object
typically of the mass of
 the large scale region.  Dynamics of under-dense
regions can be more
 complex, depending on the details of the large scale
density profile.
 Over-densities develop into virialized bound objects which
are easily
 identified in observations and N-body simulations, by means of
the
 friends-of-friends algorithm for example. The fate of under-dense
regions
 however is less certain and strongly depends on the initial
conditions.  
 Because of the complexity in studying the dynamics of matter
in
 under-dense regions, analytic solutions for special symmetric negative
perturbations are particularly important.  Most authors focused on
 symmetric
initial perturbation of the form $\delta M/M \propto r^{-\sp}$. In a matter
dominated flat universe the gravitational evolution of this profile leads to
self-similarity, which considerably simplifies the problem. Another simplifying
assumption, that is often made for evolution of under-dense system, is that of
spherical symmetry. To see the physical motivation for this assumption
consider an ellipsoidal top-hat perturbation. Since the
perturbing force (acceleration for
 negative and deceleration for positive
perturbation) is stronger along
 the minor axis, the asymmetry of a void will
decrease. Ryden (1994) has investigated the evolution of under-dense
collisionless axisymmetric systems with the initial perturbation of the form
$\delta M/M \propto r^{-\sp}f(\Theta)$, where $\Theta$ is an angle with the symmetry axis, and found that the emerging void is
nearly spherical. Fillmore \& Goldreich (1984) have found similarity
solutions for spherical perturbations in collisionless matter with $s<3$,
while Bertschinger(1985) obtained solutions for $s=3$ (uncompensated hole)
for both gas and the collisionless matter.  However, under certain
conditions, the system becomes unstable and spherical symmetry might be
destroyed. Several authors have developed different approximations to
determine the conditions for instability and to evaluate the sizes of the
formed objects (Bertschinger 1983; Vishniac 1983; Hwang,Vishniac and Shapiro
1989; White and Ostriker 1990). In the special case of a compensated void
(i.e. void surrounded by an over-dense shell whose mass is equal to the mass
deficit in the void), the dense shell typically fragments into 
large  massive clumps  (White and Ostriker 1990).
In this paper we investigate the self-similar evolution of negative perturbations for all positive values of
$s$. Our main focus is
 spherically symmetric perturbations, but, in order to
see how the
 evolution depends on geometry, we also present the main features
of
 planar and cylindrical perturbations. 
 
 The paper is outlined as
follows.  In \S 2 we write the equations of
 motion and transform them into
the appropriate self-similar form.  In
 \S 3 we derive the asymptotic behavior
of the equations for various
 values of $s$, and the adiabatic indexes, $\gamma$.
In \S 4 the
 qualitative features of the numerical solution are
described. In \S5 we check the possibility of structure formation in collisionless and baryonic shells. In \S 6 we consider evolution of shells, whose expansion is dominated by an internal pressure source. In \S 7 we present the summary of our results.

\section{The equations}
\label{equations}

We write the Newtonian equations of motion governing the adiabatic
evolution of symmetric perturbations in a collisional fluid (gas)
of adiabatic index $\gamma$. We assume that the expansion scale
factor of the universe is $a(t)\propto t^{2/3}$, where $t$ is the age of the Universe, the Hubble
function is $H(t)=2/(3t)$, the total background density is
$\rho_c=3H^2/(8\pi G)=1/(6\pi G t^2)$, and the ratio of the mean collisional
baryonic density to the critical density $\rho_c$ is denoted by 
$\Omega_{\rm b}$. 
Although self-similar evolution exists for any $\Omega_{\rm b}\le 1$,
in this paper we restrict the analysis to 
either $\Omega_{\rm b}=1$, or $\Omega_{\rm b}\ll 1$.

Denote by $r$ and $\upsilon\equiv \dd r/ \dd t$ the physical position
and velocity of a gas shell, where $ r=0$ is the symmetry center of
the perturbation.  Further, let $\rho(r,t)$ and $p(r,t)$ be the gas
density and pressure at $r$.  We write the mass within a distance $r$
from the symmetry center as $m(r,t)=\int_0^r x^{\nn-1}\rho(x,t)\dd x$,
where $\nn=1,2$, and 3 refer, respectively, to planar, cylindrical,
and spherical perturbations.  The mass within a fixed shell varies
with time like $m \sim t^{-2(3-\nn)/3}$, because of the Hubble
expansion along $3-\nn$ of the axes.  At this stage we neglect such physical effects as viscosity, thermal conduction, cooling etc., so the equations
of motion of collisional fluid are, the continuity equation,

\def\ff{\frac{2(3-\nn)}{3}}
\begin{equation}
\frac{\dd\left[\rho t^{\ff}\right]}{\dd t}=-t^\ff \rho r^{1-\nn}\partial_r(r^{\nn-1}\upsilon) \; ,
\label{eom1}
\end{equation}
the  Euler equation,
\begin{equation}
\frac{\dd \upsilon}{\dd t}-\frac{2}{9}\frac{3-\nn}{\nn}\frac{r}{t^2}
=-\frac{\partial_r p}{\rho}- \frac{4\pi G m_x}{r^{n-1}}
\; ,
\label{eom2}
\end{equation}
the  adiabatic condition,

\begin{equation}
\frac{\dd}{\dd t}(p\rho^{-\gamma})=0 \; ,
\label{eom3}
\end{equation}
and the relation,
\begin{equation}
\partial_r m= r^{\nn -1}\rho \; .
\label{eom4}
\end{equation}

In equation (\ref{eom2}), $m_x$ stands for total (collisional and
collisionless) mass.

There are several similar techniques, which we do not describe here in
detail, for calculating the collisionless mass profile
(Bertschinger 1985; Fillmore \& Goldreich 1984). The main idea is
that, since the motion is self-similar, knowing the orbit of a single
particle one can calculate the mass profile and vice versa. Given the
initial guess for mass profile, after iterative calculation of mass
profile and orbit one arrives to the self-similar solution.

The initial conditions leading to self-similar expansion are specified
at an early time close to zero, $t_i$, as
\begin{eqnarray}
\label{inid}
\frac{\delta M}{M}&=&-\left(\frac{r}{r_0}\right)^{-\sp} \; ,r\gg r_0 \\
\label{inivel} \upsilon(r,t_i)&=&\frac{2}{3t_i}r \; , \\
\label{inip} p(r,t_i)&=&0 \; ,
\label{inic}
\end{eqnarray}
where $s>0$ and $\delta M/M$ is the mean density contrast interior to
$r$. 
The Einstein-De Sitter Universe and 
the initial conditions are
completely scale-free.  The only characteristic length in the
evolution of the perturbation is the scale of non-linearity $r_*(t)$.
For self-similar collapse of positive perturbations the turn-around
radius is usually used as the scale of non-linearity. For a negative
perturbation, $r_*$ can be defined at any time as the radius interior
to which the mean density contrast has a certain fixed value.  This
means that
\begin{equation}
r_* \propto t^\alpha \qquad , \qquad
\alpha=\frac{2s}{3(s+1)}\; .
\label{ralpha}
\end{equation}
This is also the way the turnaround radius in positive perturbation
depends on time. The choice of proportionally factor in (\ref{ralpha})
is arbitrary. Here it is chosen such that a particle with initial
radius $r_i$ reach $r_*$ at $(6\delta M/M)^{-3/2}t_i$, where
the factor of 6 is arbitrary. For this choice of $r_*$ the mean
density contrast interior to $r_*$ is approximately $-0.09$ in all
symmetries.

A consequence of  self-similarity is  that 
the partial differential equations of motion can be transformed into
ordinary differential equations.
This is done by working with  $\lambda\equiv r/r_*$
and the dimensionless fluid variables,

\begin{eqnarray}
\label{scalev}\upsilon(r,t)&=&\frac{r_*}{t}V(\lambda)\\
\label{scaled}\rho(r,t)&=&\Omega_{\rm b}\rho_c D(\lambda)\\
\label{scalep}p(r,t)&=&\Omega_{\rm b}\rho_c\left(\frac{r_*}{t}\right)^2 P(\lambda)\\
\label{scalem}m(r,t)&=&\frac{1}{3}\Omega_{\rm b}\rho_c r_*^\nn M(\lambda) \; .
\end{eqnarray}

 Expressed in terms of these variables, the equations
(\ref{eom1}-\ref{eom4}) become, respectively,
\begin{equation}
\label{a1}
\left(V-\alpha\lambda\right)D'+\left(\frac{\nn-1}{\lambda}V+V'-
\frac {2\nn}{3}\right)D=0 \; ,
\end{equation}
\begin{equation}
\label{a2}\left(\alpha-1\right)V+\left(V-\alpha\lambda\right)V'
-\frac{2}{9}\frac{3-\nn}{\nn}\lambda
=
-\frac{P'}{D}-\frac{2}{9}\frac{M}{\lambda^{\nn-1}} \; ,
\end{equation}

\begin{equation}
\label{a3}\left(\gamma\frac{D'}{D}-
\frac{P'}{P}\right)\left(V-\alpha\lambda\right)=2\left(\alpha-2+\gamma\right)
\; ,
\end{equation}

\begin{equation}
\label{a4}M'=3\lambda^{\nn-1}D \; ,
\end{equation}
where the prime symbol denotes derivatives with respect to $\lambda$.

Self-similarity implies that the shock appears (if it does) at fixed
$\lambda=\lambda_{\rm s}=r_{\rm s}/r_*$, so the physical radius of the shock
$r_{\rm s}\propto t^\alpha$ and its non-dimensional speed is $(r_*/t)^{-1}(\dd
r_{\rm s}/\dd t)= \alpha\lambda_{\rm s}$.  At the surface of the shock the fluid
variables satisfy the jump conditions obtained from mass, momentum,
and energy conservation (cf. Spitzer 1978, \S 10.2):

\begin{eqnarray}
\frac{\upsilon'_+}{\upsilon'_-}&=&\frac{\rho_-}{\rho_+}=\frac{\gamma-1}{\gamma+1},\\
p_++\rho_+\upsilon_+^{'2}&=&p_-+\rho_-\upsilon_-^{'2},\\
\upsilon_-&=&\upsilon'_-+\frac{dr_{\rm s}}{dt},\\
\upsilon_+&=&\upsilon'_++\frac{dr_{\rm s}}{dt},
\end{eqnarray}
where the superscripts of the minus and plus signs refer to pre- and
post-shock quantities and $\upsilon'$ is the velocity relatively to the shock position.

In terms of the non-dimensional fluid
variables we obtain
\begin{eqnarray}
\label{jump1}
V^+&=&\alpha\lambda_{\rm s}+\frac{\gamma-1}{\gamma+1}(V^- -\alpha\lambda_{\rm s})\; ,\\
\label{jump2}D^+&=&\frac{\gamma+1}{\gamma-1}D^- \; ,\\
\label{jump3}P^+&=&\frac{2}{\gamma+1}D^-(V^--\alpha\lambda_{\rm s})^2\; .
\end{eqnarray}
Outside the shock, at $\lambda >\lambda_{\rm s}$, the pressure vanishes and
the pre-shock fluid variables can be found by solving the equations
(\ref{a1}-\ref{a4}) with zero pressure. Analytic solutions for zero
pressure exist for planar and spherical geometries, but not for
cylindrical (Zel'dovich 1970, Peebles 1980, Bertschinger 1985).  

\subsection{The existence of shocks and the value of $s$}
Lower and upper limits on $s$ exist for a shocked expansion of an
isolated negative cosmological perturbation.  A necessary condition
for the formation of a shock is that the perturbation is steep enough
so that inner shells expand faster than outer shells.  An inspection
of the equations of motion shows that this happens only if
$(\alpha-1)\alpha< \frac{2}{9}\frac{3-n}{n}$, that is $s>2$ for
spherical, $s>1.54$ for cylindrical, and $s>1$ for planar symmetry.
This lower limit disappears if an external energy source is placed at
the center of the perturbation. 

 An upper limit on $s$ also exists. To
see this, consider the case of a spherical perturbation.  If the
perturbation is negative, the total energy of a particle is positive
and so the energy inside the shock must increase as it sweeps more
particles.  On the other hand, self-similarity implies that the total
energy within $\lambda_{\rm s}$ varies with time as $t^{5\alpha-4}$ which is
a decreasing function of time for $\alpha < 4/5$, i.e., $s< 5$. To
satisfy both conditions the energy within $\lambda_{\rm s}$ must be
negative. The only way to construct such system is by placing a bound
object (i.e., negative energy) at the center. However, since there is no 
way to make this
object grow with time, the evolution of the perturbation never
becomes self-similar.
The limiting case of
$s\rightarrow 5$, which corresponds to a  shock in a homogeneous
medium (compensated void), was investigated by Bertschinger (1983).
The corresponding upper limits for cylindrical and planar perturbations
are  $s\approx 3.5$ and $s=2$, respectively.

\subsection{Boundary conditions}
Solving the self-similar equations (\ref{a1}--\ref{a4}) requires
boundary conditions on the fluid variables.  We will see in the next
section that shocked expansion is associated with a dense shell whose outer
boundary is the shock. 
Outside the shock the pressure is zero and the
equations can be solved by standard numerical techniques. The jump
conditions (\ref{jump1}--\ref{jump3}) provide  post-shock fluid variables.
Given the  post-shock fluid variables, the
equations (\ref{a1}--\ref{a4}) can be integrated inward to obtain the 
fluid variables inside the shock.

So far it seems that one can introduce the shock at any arbitrary
radius. However, inward integration of the equations from an assumed shock
position always end at a singular point, $\lambda_0$, where at
 least one of
the fluid variables becomes infinite. Since there is no
 natural way of
eliminating this singularity, this must be the inner
 boundary of the baryonic
shell, $M_{gas}(\lambda_0)=0$. If the total energy of the system is to
be conserved, the pressure at the inner boundary of the system must be zero,
$P(\lambda_0)=0$,
 otherwise, energy will continuously be injected into the
system. It turns out that there is at most one value of $\lambda_{\rm s}$, for
which
 the solution satisfies $P(\lambda_0)=0$.  Smaller values of
$\lambda_{\rm s}$ give positive mass and pressure at $\lambda_0$, while larger
values give zero mass and positive pressure.
 
\section{Asymptotic behavior near the inner boundary of the shell}
\subsection{Baryonic matter}

To obtain an asymptotic behavior for the collisional matter we expand
the fluid variables near $\lambda_0$, as
\begin{eqnarray*}
M(\lambda)&=&M_0(\lambda-\lambda_0)^\mu \\
D(\lambda)&=&D_0(\lambda-\lambda_0)^\delta \\
P(\lambda)&=&P_0(\lambda-\lambda_0)^\eta\\
V(\lambda)&=&\alpha\lambda_0+V_0(\lambda-\lambda_0)^\nu
\end{eqnarray*}

By substituting these expressions in the equations of motion we find
that $\nu=1$ in all cases(cf. Ostriker \& McKee 1988), while
$\delta$, $\mu$ and $\eta$ depend on $s$ and $\gamma$. 
The asymptotic exponents of the fluid variables in the mixed and
purely collisional matter are the same. This differs from the collapse
case where the collisional component can change the asymptotic
exponents of the collisional fluid variables near the center (Chuzhoy
\& Nusser 2000).
Table 1 lists
the asymptotic exponents  for the following three cases:

\noindent {\it I. Expansion without a shock, $\lambda_0=0$:}
As we have mentioned in the previous section, there are no solutions with a shock
for $(\alpha-1)\alpha\ge \frac{2}{9}\frac{3-n}{n}$ with
$P(\lambda_0)=0 $.  The expansion then develops without a shock and
the pressure vanishes everywhere.  Substituting the asymptotic form
for the fluid variable with $\lambda_0=0$, in the equation of motion
yields the asymptotic exponents in this case as listed in Table 1.
Near $\lambda=0$, the mass vanishes and the velocity grows linearly as
$V_0\lambda$, where $V_0$ is independent of $s$.

\noindent {\it II. Shocked expansion with $P(\lambda_0)=0$:}
This occurs if $(\alpha-1)\alpha< \frac{2}{9}\frac{3-n}{n}$, i.e., if
$s>2$ for spherical, $s>1.54$ for cylindrical and $s>1$.  For these
values of $s$ the exponent $\delta$ as given in table 1 is always
negative and so the density diverges at $\lambda_0$.

\noindent {\it III. Shocked expansion with $P(\lambda_0)>0$:}
This occurs when  energy can be injected from the center of the
perturbation (e.g., SN energy injection) allowing
$P(\lambda_0)=const>0$  at the inner shell boundary.
The asymptotic behavior for this
case is also listed in the table.

\subsection{Collisionless matter}

After going through  first shell-crossing, a collisionless particle
 starts its
converging oscillatory motion around some $\lambda_x$ inside the dense
shell (Fillmore \& Goldreich 1984). At each turnaround point there is
a caustic peak, where the density diverges. To obtain the asymptotic
solution at the caustic we again expand the variables around the peak
location
\begin{eqnarray*}
M(\lambda)&=&M_0(\lambda-\lambda_0)^\mu \\
D(\lambda)&=&D_0(\lambda-\lambda_0)^\delta \\
P(\lambda)&=&0\\
V(\lambda)&=&\alpha\lambda_0+V_0(\lambda-\lambda_0)^\nu \; .
\end{eqnarray*} 
Substituting into the continuity, Euler and mass equations, we obtain
$\mu=1/2$, $\delta=-1/2$, $\nu=1/2$ and
$V_0^2=2(1-\alpha)\alpha\lambda_0$. 
Negative and positive  $V_0$ correspond to the 
incoming and out-coming streams at the caustic. 
Note that, unlike the collisional case, the asymptotic
exponents depend neither on symmetry nor value of $s$.

\begin{table}
\caption{ Asymptotic constants, $V_0, \delta,\eta$ collisional
fluid variables in a flat universe with no collisionless matter. }
\begin{tabular}{l|c|c|c|}\hline
&$\lambda_0=0$ &$ \lambda_0>0$& $\lambda_0>0$ \\
& & &  \\
&NO  SHOCK & $P(\lambda_0)=0 $& $P(\lambda_0)>0$\\  \hline
& & &  \\
$\delta $&$\frac{ns(\sqrt{n}-\sqrt{n+24})}{s\sqrt{n+24}-(s+4)\sqrt{n}}$&
$\frac{n-2-s(3\gamma-4)}{2+n(\gamma-1)+s(3\gamma-4)} $&
 $-\frac{s(3\gamma-4)+2}{2+n\gamma+s(3\gamma-4)} $\\
& & &  \\
\hline
& & &  \\
$\mu$ & $\delta+3 $& $\delta+1 $& $\delta+1$\\
& & &  \\   \hline
& & &  \\
$\eta$ & $P=0 $& $\delta+1 $& \\
& & &  \\  \hline
& & &  \\
$V_0$ & $\frac{3\sqrt{n}+\sqrt{n+24}}{6\sqrt{n}}$& $\frac{2(n-2+\gamma (1-n)+2s(\gamma-2))}{3\gamma s}$& \\
& & &  \\   \hline
\end{tabular}
\label{tab:3}
\end{table}

\section{Numerical solutions}

In order to obtain the detailed structure of the baryonic shell we
numerically integrate the self-similar equations of motion
(\ref{a1}-\ref{a4}). The location of the shock, $\lambda_{\rm s}$,
is sought by demanding vanishing pressure at the inner boundary of the
baryonic shell.  Integrating the equations inwards from arbitrary
$\lambda_{\rm s}$ always leads to a singular point, where the
derivatives of the fluid variables diverge.  If the assumed
$\lambda_s$ is larger than the desired value, then at this point
pressure is positive and mass is zero, while if it is smaller then the
mass is also positive. Thus it is possible to find $\lambda_{\rm s}$
iteratively starting from the initial guess and varying it in
accordance with the results of the integration.  We will  show
numerical results for only two cases $\Omega_{\rm b}=1$ and
$\Omega_{\rm b}\ll 1$. To find the solution for $\Omega_{\rm b}\ll 1$
we need the mass profile for the purely collisionless collapse. This
is found using the method of Fillmore\& Goldreich(1984) and
Bertschinger(1985).

\begin{figure}
\centering
\mbox{\psfig{figure=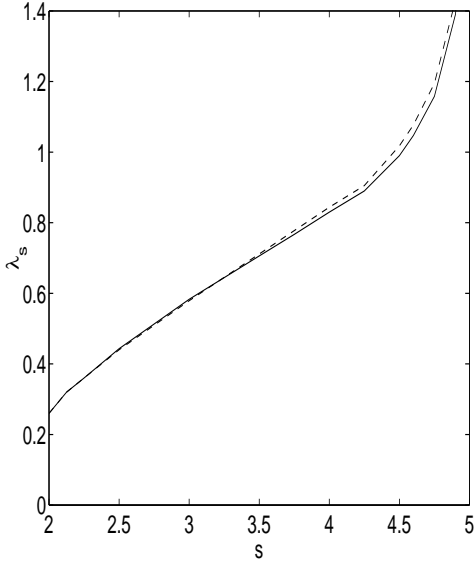,height=3.0in,width=2.5in}}
\caption{ The shock location $\lambda_{\rm s}$ as a
function of $s$. The dashed and the solid line correspond, respectively to $\Omega_{\rm b}=1$ and $\Omega_{\rm b}\ll1$. } \label{lam}
\end{figure}

\begin{figure}
\centering
\mbox{\psfig{figure=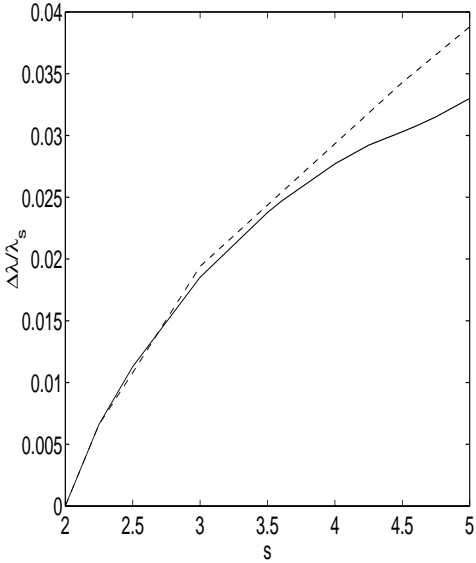,height=3.0in,width=2.5in}}
\caption{
  Shell width $\Delta\lambda/\lambda_{\rm s}$ as a function of $s$.
  The solid and the dashed line correspond, respectively to
  $\Omega_{\rm b}=1$ and $\Omega_{\rm b}\ll1$.  }
\label{dlam}
\end{figure}

\begin{figure}
\centering
\mbox{\psfig{figure=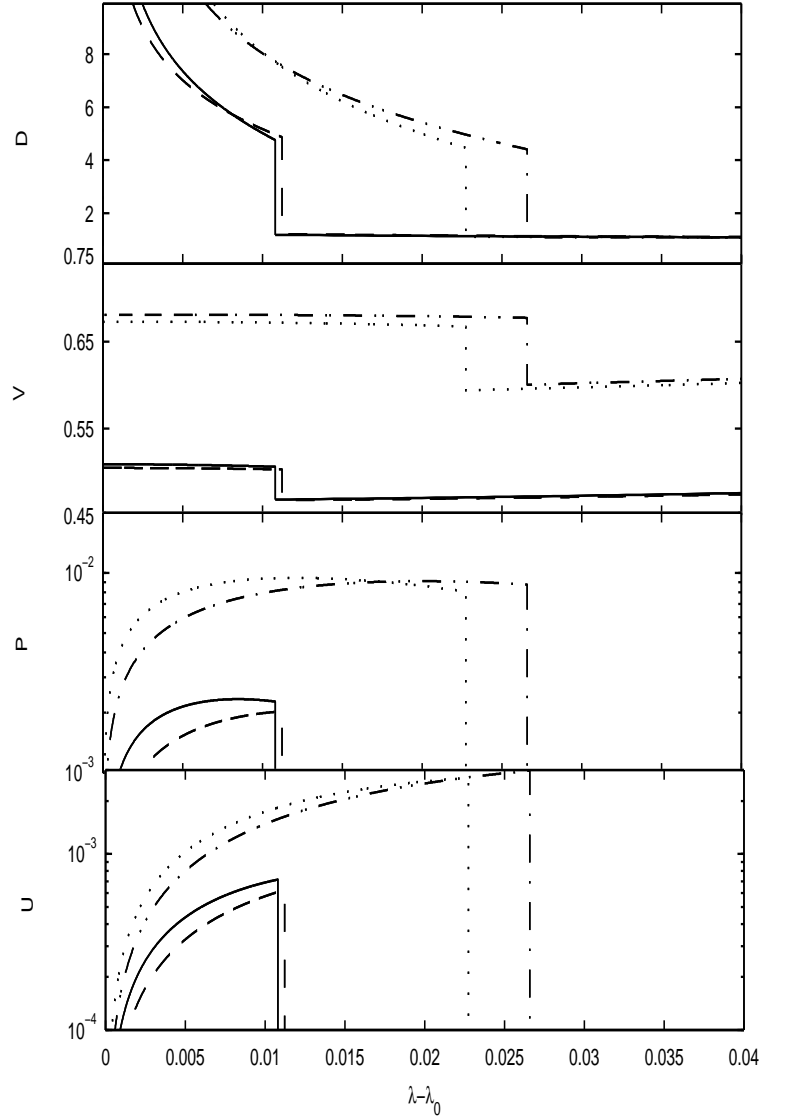,height=6.0in,width=4.0in}}
\caption{ The fluid variables (density, velocity, pressure and
thermal energy) for the case of spherical symmetry and
$\gamma=5/3$.  The solid and dotted lines correspond to $\Omega_{\rm b}=1$ with  $s=3$ and $s=4$ respectively , dashed and dash-dotted lines
  correspond to $\Omega_{\rm b} \ll 1$ with $s=3$ and $s=4$.}
\label{varf}
\end{figure}

Figures \ref{lam}-\ref{dlam} 
show the shock location and thickness as a function of $s$. Figure
\ref{varf} shows the fluid variables as a function of $\lambda$ for
several values of $\sp$.  All the plots were made for a spherical
expansion with $\gamma=5/3$.  All the numerical solutions we found
conform near the inner boundary of the shell to the results of
asymptotic analysis.  The location of the shock (fig. \ref{lam})
depends strongly on $s$ but is nearly independent of both of $\gamma$ and
$\Omega_{\rm b}$. The reason for this weak dependence on $\gamma$ is
that the thermal energy is small to the kinetic and gravitational
energies. The thermal energy and hence the width of the shell increase
with $s$ (fig. \ref{dlam}), from zero to a few per cent of the void
radius.  The baryonic shell lies in the collisionless shell.
Because the collisionless particles are not slowed down as they enter
the shell, the collisionless shell is typically several times wider
than the baryonic shell. 

\section{Fragmentation}
\subsection{Baryonic matter}

We address the question whether the shells can fragment into bound objects, such as galaxies. 

For the expansion of an object to halt, its gravitational energy must be larger than the sum of the thermal and the kinetic energy by some finite factor, i.e.

\begin{eqnarray}
\label{freq}E_{\rm g}+C_f(E_{\rm k}+E_{\rm t})<0 \; 
\end{eqnarray}

To check whether an object satisfying the above equation can be found, we calculate the energy of a disk placed on the interior boundary of the shell (where the density is highest and the temperature is lowest).

 Let the disk be of radius $l=r_* L$ and height $ h=r_*
H$. 

The kinetic energy due to the shear velocity is found to be
\begin{equation}
E_{\rm k}=C_n m_{\rm d}\left(\frac{l}{2t}\right)^2= 
(\frac{r_*^5}{6t^4}) 
\Omega_{\rm b} M
L^4\frac{C_n}{12\lambda_0^2} \; ,
\end{equation}

where $C_n=4/9$, $(2/9+\alpha^2/2)$, and $\alpha^2$ for $n=1,2$, and 3,
respectively and $m_{\rm d}\approx\frac{\pi l^2}{3r_{\rm s}^2}M\Omega_{\rm b} \rho_c r_*^3$ is the mass of the disk.

The gravitational energy is obtained using the thin shell approximation 
\footnote{For all the disks satisfying (\ref{freq}), that we found, $H\ll L$. Thus, since the correction for finite thickness of the disk lowers the gravitational energy, the assumption of a thin disk is justified.} -

\begin{eqnarray}
E_{\rm g}=-\frac{8m_{\rm d}^2}{3\pi l}=-(\frac{r_*^5}{6 
t^4})\frac{4}{81\pi}\Omega_{\rm b}^2L^3\frac{M^2(\lambda_0+H)}{\lambda_0^4}
\end{eqnarray}

The thermal energy -

\begin{eqnarray}
E_{\rm t}=\int\frac{\pi}{\gamma-1}l^2p dr=(\frac{r_*^5}{6
t^4})\frac{\Omega_{\rm b}}{\gamma-1} L^2\int_{\lambda_0}^{\lambda_0+H
}Pd\lambda \; 
\end{eqnarray}

The kinetic energy due to the radial velocity -

 \begin{eqnarray}
E_{\rm k2}=\int\pi l^2 \rho(v-<v>)^2/2 dr= \\
=(\frac{r_*^5}{6
t^4})\Omega_{\rm b} L^2\int_{\lambda_0}^{\lambda_0+H
}D(V-<V>)^2/2d\lambda \; ,
\end{eqnarray}
where $<V>$ is an average radial velocity of the disk.
From the asymptotic analysis follows that $E_{\rm k2}/E_{\rm t}\propto H$, so for $H\ll 1$ we can neglect  $E_{\rm k2}$.

For large values of $L$ the kinetic energy term, $E_{\rm k}\propto L^4$,
would be dominant, while for small $ L$ the thermal energy, $E_{\rm
  t}\propto { L}^2$, is the only significant term. Thus the disk can be
unstable only for intermediate values of $ L$, where the gravitational
term, $E_{\rm g}\propto L^3$, is dominant.  From eq. (\ref{freq}) we can
factor an equation of second degree in $ L$, whose roots determine the
limits of instability for a given thickness of the disk, $H$.  The
condition for the existence of the roots is
\begin{equation}
\label{ineq}
16\Omega_{\rm b}(\gamma-1)\frac{M^3}{\lambda_0^6}\;>\;3^7\pi^2 C_n
C_f^2\int_{\lambda_0}^{\lambda_0+H}Pd\lambda
\end{equation}

\subsubsection{Fragmentation without cooling}

The asymptotic solution implies that the term on the l.h.s. of eq. \ref{ineq} grows as
$H^{3(\delta+1)}$ and the right as $H^{\delta+2}$ and thus in the
limit $H\rightarrow 0$ 
the inequality is satisfied if $\delta<-\frac{1}{2}$ .

For $\gamma=5/3$, $\delta$ is less than $-\frac{1}{2}$ for all planar shocks and for cylindrical shocks with $s>10/3$, which are consequently gravitationally unstable even for $\Omega_{\rm b}\ll 1$. For spherical shocks $\delta$ is less than $-\frac{1}{2}$ only in case the shell pushed by an internal pressure source with expansion rate $\alpha<8/9$.
 For spherical shocks with $P(\lambda_0)=0$  the numerical integration shows that for $\Omega_{\rm b}=1$ and $C_f=1$ eq. (\ref{ineq}) is satisfied only for $s>\sim3.5$ and not at all if $\Omega_{\rm b}<0.8$.

\subsubsection{Fragmentation with cooling}
From the numerical calculation we find that for a disk on a spherical shell with zero internal pressure to
satisfy eq.(\ref{ineq}) its thermal energy must fall at least by factor 
$\Omega_{\rm b}^{-1}$. To check when this is indeed possible we must introduce  a physical scale into the problem. 

The physical temperature of the shock for ionized gas is
\begin{eqnarray}
\label{ptemp}
T\approx\frac{P_{\rm s}}{D_{\rm s}\lambda_{\rm s}^2}\frac{0.6m_p}{k_B}\left(\frac{r_{\rm s} }{t}\right)^2
\end{eqnarray}

\begin{figure}

\centering

\mbox{\psfig{figure=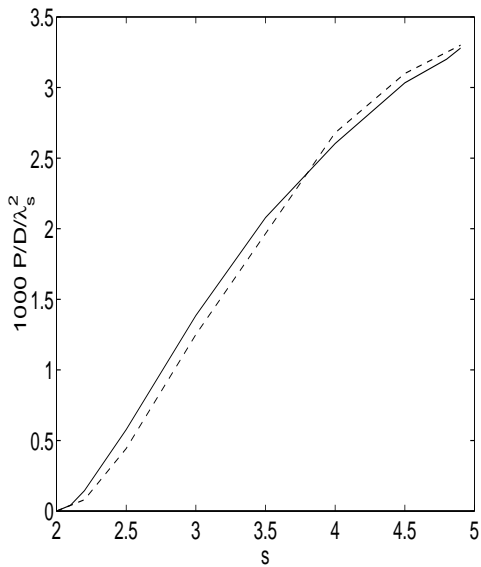,height=3.0in,width=2.5in}}
\caption{ Non-dimensional shock temperature as a function of $s$
 (eq. \ref{ptemp}). The solid and the dashed line correspond, respectively, to
 $\Omega_{\rm b}=1$ and $\Omega_{\rm b}\ll 1$. 
 }
\label{avg}
\end{figure}

 For spherical symmetry the value of $P_{\rm s}/D_{\rm s}\lambda_{\rm s}^2$  can be approximated by
 $(s-2)/10^3$ for shell driven by negative density perturbation  (fig. \ref{avg}) or by $3(\alpha-2/3)^2/16$ for shell driven by internal pressure. Thus

\begin{equation}
\label{Tu1}
 T\approx 2(s-2)10^5 r_{10}^2(1+z)^{(s-2)/s}K 
\end{equation}

or

\begin{equation}
\label{Tu2}
T\approx 4(\alpha-2/3)^2 10^7 r_{10}^2(1+z)^{3(1-\alpha)}K ,
\end{equation}

where $r_{s0}=10 r_{10}$ \rm {Mpc} is the present radius of the shock.

At temperatures higher than $\sim 10^4 K$ and densities not much higher than the background values the inverse Compton scattering is  the dominant cooling process. At redshifts $\gtsim 8$ the cooling time  $t_{cool}=1.6\cdot
10^{12}(1+z)^{-4}\rm Yr$ (Bertschinger 1983) is less than the age of the universe. This means that the gas shock cools quickly after the shock and the average temperature of the shell remains at constant value of order of $10^4 K$.
Thus the approximate condition for fragmentation is for the shock temperature to be higher than $\sim 10^4\Omega_{\rm b}^{-1} K$.

When the collapse depends on cooling the mass of the formed objects is of order of Jeans mass at $T\sim 10^4 K$, i.e.,  $M\sim 10^{11}\Omega_{\rm b}^{-1/2}(1+z)^{-3/2}M_\odot$. Since the mass accretion rate on the shell is proportional to $(1+z)^{9(1-\alpha)/2}$, the number of objects of the mass $M$ formed by the shell is proportional to $N(M)\propto M^{2\alpha-3}$.

It should be noted that different equations of motion (Appendix A) are needed to describe the shell that loses its thermal energy during fraction of its dynamic time (snow-plow). The main effect of the cooling is to move the shock inwards thus decreasing the swept mass and the shock temperature.
However, since in this problem the thermal energy constitutes only a small fraction of the total, these differencies are generally insignificant.

\subsection{Collisionless matter}
We now address the stability condition of the collisionless shell. The kinetic
and gravitational energy terms are the same as in the baryonic case. 
There are two stream flows near  $\lambda_0$ (cf, section 3.2), so 
 the ``thermal'' energy resulting from the velocity dispersion in the flow is
$E_{\rm t} =\frac{r_*^5}{6\pi
 t^4}\frac{\pi}{2}
L^2\int_{\lambda_0}^{\lambda_0+H}D(V-\alpha\lambda)^2d\lambda$. The analog of
eq. \ref{ineq} for this case is
\begin{equation}
\label{ineq2}
{32}\frac{M^3}{\lambda^6}>3^7 \pi^2 C_f^2
C_n\int_{\lambda_0}^{\lambda_0+H}D(V-\alpha\lambda)^2d
\lambda
\end{equation}
Near the inner boundary both the left and the right term grow as $H^{3/2}$
and thus the fact whether the last inequality is satisfied depends on the
choice of $C_f$. However, using just the lower constraint $C_f>1$, we found
that for small $H$ it is never satisfied in the case of spherical symmetry. For large
$H$ the energy analyses must give the same results for pure
collisional ($\Omega_{\rm b}=1$ and $\Omega_{\rm c}=0$) and pure collisionless ($\Omega_b=0$ and $\Omega_{\rm c}=1$) models. Thus we expect the bound
 structures to form for all $s$ in planar shells, for $s>\sim 3$ in cylindrical shells, and possibly for $s>\sim 3.5$ in spherical shells.


\section{Explosive amplification}
The explosive amplification model suggesting that galaxies can
originate from series of explosions was proposed by Ostriker\&
Cowie(1981) and later developed by other authors. Vishniac, Ostriker
\& Bertschinger (1985) showed that an energy injection from supernova
explosions in the range $10^{57}-10^{61}$ could produce a void
reaching $\sim 10 {\rm Mpc}$. The cooling of the shell surrounding the
void would then lead to the formation of galaxies. However, since the
cooling proceeds chiefly through inverse Compton scattering, it was
argued that a large number of such explosions would be incompatible
with the observational limits on the Compton y-distortion parameter
(Levin, Freese \& Spergel 1992). Later Miranda\& Opher(1997) showed
that if the energy was injected in several rather than in a single
cycle, then the distortion would be several times smaller. In their
calculation the current observational limit $y\lsim 1.5\cdot 10^{-5}$
(Fixen et al. 1996) is compatible with voids of present radii $R\sim
10-30$ {\rm Mpc}, the filling factor $f_F\sim 0.07-0.30$, and $\Omega_{\rm
b}=0.1$. In this section we explain how the distortion produced by
shocks of similar size might be still lower by one or two orders of
magnitude.

In the previous section we have shown that before $z\sim 8-10$ large
baryonic shells become gravitationally unstable. The collapse of one
object creates a pressure gradient causing the surrounding gas to
expand and thus creates a feedback against further collapse. Therefore
we expect that fragmentation proceeds continuously rather than in
cycles. Depending on the star formation efficiency in the collapsed
objects and the fraction of stars turned to supernovae (SN), the
dynamics of the shell might become dominated by the energy input from
the SN explosions. By using the approximation $L_{inj}\propto
\dot{m_s} t^q$, where $L_{inj}$ is the energy injection rate and
$\dot{m_s}$ is the mass accretion rate on the shell, we get the
expansion rate of the shell $r_s\propto t^\alpha$, where
$\alpha=1+q/2$. Note that if mass to energy conversion
efficiency doesn't change drastically over time, then $q\approx 0$ and
$\alpha\approx 1$.

If the shell expansion is dominated by internal pressure then the collapsing objects, whose surface area is decreasing,  would be eventually overtaken by the shell and, unless $\Omega_{\rm b}\sim 1$, remain inside.  Since the interior
of the shell is empty of gas, only a small fraction of the SN energy
is turned into a thermal energy, making the CBR distortion in
this scenario much smaller than in previous models (see Appendix
B). Similarly, because less energy is lost through cooling, larger
voids can be produced for the same mass to energy conversion
efficiency.  In this case the radius of the baryonic shell is
$R\approx 10 h^{-1} (\frac{\epsilon}{10^{-5}})^{1/2}(1+z)^{-3/2}$
\rm {Mpc}, where $\epsilon={E_{inj}}/{m_{shell}c^2}$.

\section{Summary}
 
For the gaussian initial density field with scale free power spectrum the rms density perturbation is $<(\frac{\delta M}{M})^2>\propto r^{-(n+3)}$, where $-3<n<4$. It is therefore reasonable to identify  $s=(n+3)/2$, from which follows $0<s<3.5$. Thus in case $n>1$ we might expect that the universe contains a large number of spherical shells. We have shown in the paper that the evolution of spherical perturbations with $s>2$ results in formation of overdense regions, which, unless $(s-2)\ll 1$ may contain a large fraction of matter.

A distinctive property of structure formation in the shell is the segregation between the baryons and the collisionless matter. The energy analysis shows that for $s<3.5$ (i.e. $n<4$), which follows from the gaussian density field, the collisionless matter cannot form bound objects, while at high redshifts cooling allows formation of the baryonic objects on the appropriate scales. Similarly, the explosive model (\S 6) predicts that the baryonic shell is driven by pressure ahead of the collisionless and fragments in an independent way. This is entirely different from the evolution of positive perturbations, where gas must fall into the potential well formed by collisionless matter.

The energy analysis (\S 5.1.2) shows that the mass of the bound objects formed from the baryonic shell is of order of Jeans mass at $T\sim 10^4 K$ -   
$M\sim 10^{11}\Omega_{\rm b}^{-1/2}(1+z)^{-3/2}M_\odot$.
This is again different from the evolution of positive perturbation, which predicts the Jeans mass to be also dependent on the spectrum of perturbations.

An observational limit on the scale and number of the shells can be obtained from the amplitude of Compton y-distortion. 
Both for the shell driven by the negative density perturbation and for the shell driven by internal pressure the predicted amplitude is of order $10^{-5}\frac{\Omega_{\rm b}}{0.1}(\frac{r_{10}}{3})^2f_F$, where $f_F$ is the present filling factor  (see Appendix B for the details of the calculation).

The current upper limit  $y=1.5\cdot 10^{-5}$ is thus consistent with shells, whose present radius is below $\sim 30$ \rm {Mpc} and the filling factor is of order of unity.


\section{acknowledgment}
This
research was supported by the Technion V.P.R Fund- 
Henri Gutwirth Promotion of Research Fund,  the German-Israeli Foundation
for Scientific Research and Development, and the Israeli Science Foundation.

\newpage

\appendix
\onecolumn

\section{Expansion of a thin momentum conserving shell (snow plow) }

If the cooling time is much shorter than the dynamical time of the
system the shocked gas would be compressed into a thin shell, whose
expansion can still be described in self-similar form
($r_s=r_{s0}t^\alpha$).  Given the initial density profile and energy
injection rate, the constants 
$\alpha$ and $r_{s0}$ can be determined using momentum
conservation, i.e.,
\begin{equation}
\label{mom}
\frac{d(mt^{2-2n/3\alpha r/t})}{dt}=\frac{2(3-n)}{9n}\frac{mt^{2-2n/3}r}{t^2}-\frac{2\pi (m+2m_c)m t^{2-2n/3}}{r^{n-1}}-\rho r^{n-1}(v-\alpha r/t)v t^{2-2n/3}+f_p,
\end{equation}
where $f_p$ is the force of the internal pressure, $\rho$ and $v$ are
the density and the velocity of the gas just outside the shock, and $m$
and $m_c$ are, respectively, the masses of baryonic and collisionless
components.
In the dimensionless form (\ref{mom}) is expressed as,
\begin{equation}
\label{mom1}
6\alpha(\alpha-1+n(\alpha-2/3))=\frac{2(3-n)}{n}-C\frac{M}{\lambda^n}+\frac{27DV(\alpha\lambda-V)\lambda^{n-2}}{M}+\frac{F_p}{M\lambda},
\end{equation}
where $C=2$  for
  $\Omega_{\rm b}\ll 1$ (if the baryonic shell is ahead of the collisionless) and $C=1$ for   $\Omega_{\rm b}=1$.

 If 
 the expansion is caused by instantaneous energy injection
then $\alpha$ is found by substituting $M=3\lambda/n$, $D=1$,
$V=2\lambda/3$ and $F_p=0$ in (\ref{mom1}). The result is
\begin{equation}
\label{al}
\alpha=\frac{3+4n+\sqrt{(24+25n-12C(n+1))/n}}{6(n+1)} \; .
\end{equation}

For $\Omega_{\rm b}=1$ $\alpha$ is equal respectively to 0.797 for spherical symmetry, 0.853 for cylindrical and 1 for planar symmetry. For $\Omega_{\rm b}\ll 1$ and the baryonic shell is ahead of the collisionless $\alpha\approx 2/3+\Omega_{\rm b}/2n$.

If on the other hand the expansion is caused by negative density
perturbation or a continuous energy injection then while $\alpha$ is
the same as in the adiabatic case where cooling is negligible. 
This is because the thermal energy is only a small fraction of the
total energy anyways.
Cooling however causes a slight decrease in $r_{s0}$.

\section{Shock imprint on the cosmic background radiation}

The spectral distortion of the cosmic background radiation (CBR) is given by
\begin{equation}
y=\frac{\Delta T}{2 T}=\int \frac{k_B T_e}{m_e c^2}\sigma_T n_e dl \; ,
\end{equation}
where $n_e$ and $T_e$ are the electron density and temperature and
$\sigma_T$ is the Thompson cross section for photon electron scattering.

Averaging over a spherical volume gives 
$<{dy}/{dl}>={\sigma_T E_{th}}/({4\pi m_e c^2 r_s^3})$,
where $E_{th}$ is the thermal energy of the entire shell.

Prior to $z\lsim 8$ the shell cools quickly so its thermal energy may be approximated by
\begin{equation}
E_{th}\approx\int_0^\infty \frac{6\pi\dot{m_s}}{\mu}k_B T_s
e^{-t/t_{cool}}dt=\frac{6\pi\dot{m_s}}{\mu}k_B T_s
t_{cool}=(\frac{3\alpha-2}{3}\frac{r_s^5 t_{cool}}{Gt^5})( \frac{P_s
M_s}{D_s\lambda_s^5}) \; ,
\end{equation}
where $\mu$ is the mean mass per particle.
At later redshifts the thermal energy is $E_{th}=\int_{r_0}^{r_s}4\pi p r_s^2 dr\approx (\frac{2r_s^5}{3Gt^4})(\lambda_s^{-3}\int_{\lambda_0}^{\lambda_s}P d\lambda)$.

The filling factor of the shells grows as $f(t)=f_F t^{3\alpha-2}$,
where $f_F$ is the final filling factor, thus the final distortion is,
\begin{equation}
\label{dist}
y=\int^{z_f}_{z_i} \frac{dy}{dl} f(t)\frac{dl}{dz}dz \; .
\end{equation}

Both for spherical shells driven by the internal pressure
($4/5<\alpha<2$) and shells driven by the negative density
perturbation  $y\lsim 10^{-5}f_F r_{10}^2 \Omega_{\rm b}$,
where $f_F$ and $10r_{10}$ Mpc are the present filling factor and
radius of the shells .

\end{document}